# Fault-tolerant detection of a quantum error

S. Rosenblum[1,2,†,*], P. Reinhold[1,2,†], M. Mirrahimi[2,3], Liang Jiang[1,2], L. Frunzio[1,2], R.J. Schoelkopf[1,2]

[1]Departments of Applied Physics and Physics, Yale University, New Haven, CT 06511, USA.

[2]Yale Quantum Institute, Yale University, New Haven, CT 06520, USA.

[3]QUANTIC team, INRIA de Paris, 2 Rue Simone Iff, 75012 Paris, France.

[†]These authors contributed equally to this work

[*]email: serge.rosenblum@yale.edu

**Abstract**: A critical component of any quantum error–correcting scheme is detection of errors by using an ancilla system. However, errors occurring in the ancilla can propagate onto the logical qubit, irreversibly corrupting the encoded information. We demonstrate a fault-tolerant error-detection scheme that suppresses spreading of ancilla errors by a factor of 5, while maintaining the assignment fidelity. The same method is used to prevent propagation of ancilla excitations, increasing the logical qubit dephasing time by an order of magnitude. Our approach is hardware-efficient, as it uses a single multilevel transmon ancilla and a cavity-encoded logical qubit, whose interaction is engineered in situ by using an off-resonant sideband drive. The results demonstrate that hardware-efficient approaches that exploit system-specific error models can yield advances toward fault-tolerant quantum computation.

**Main Text:**

In a fault-tolerant (FT) implementation of an error-corrected quantum circuit, the failure of a single component results in at most a correctable error in the output (*1*). Scalable quantum computation will ultimately require fault tolerance for every part of a logical circuit, including state preparation, gates, measurements, and error correction (*2*). FT error syndrome measurements are a particularly crucial element because they appear frequently in every encoded circuit. For a quantum nondemolition (QND) measurement of an error syndrome to qualify as FT, it must go beyond just leaving the syndrome unchanged, and also protect the logical degrees of freedom.

Typically, non-fault-tolerance in a syndrome measurement arises from errors in the ancilla that propagate to the logical qubit, where they can cause uncorrectable errors. A common proposed strategy is to introduce multiple ancillae, each interacting with a restricted number of physical qubits that make up a single logical qubit (*3, 4, 5, 6*). While this may prevent ancilla errors from spreading in the system, it comes at the cost of an increased hardware overhead. Alternatively, in the Bacon-Shor subsystem approach (*7*) ancilla errors are allowed to accumulate in gauge degrees of freedom, which need not be monitored or corrected. Recently, this type of syndrome measurement was demonstrated in both trapped ions and superconducting qubits using a four-qubit code which allows error detection, but not error correction (*8, 9*).

In this work, we implement a FT error syndrome measurement using a different approach. In our method, symmetries present in the system-ancilla interaction make it invariant under the action of dominant ancilla errors, preventing their propagation to the system in any form. In other words, the system-ancilla interaction is designed to commute with the dominant error operators, and thus errors occurring during the interaction are equivalent to errors occurring afterwards. This form of protection, called error-transparency (*10*), extends concepts related to decoherence free subspaces (*11*) in order to realize FT operations.

We implement our FT syndrome measurement on a logical qubit encoded in a single 3D superconducting cavity ($\omega_c / 2\pi$ = 4.5 GHz, $T_1^c$ = 1.1 ms). We encode quantum information using the Schrödinger cat code (*12, 13, 14*), whose codewords $|0/1\rangle_L = |C^+_{\alpha/i\alpha}\rangle$ are superpositions $|C^+_\alpha\rangle \propto |\alpha\rangle + |-\alpha\rangle$ of coherent states, where we use mean photon number $|\alpha|^2 = 2$. The dominant cavity error, single-photon loss, causes the photon number parity of both codewords to change from even to odd, without destroying the encoded information. Parity is therefore the error syndrome, and the information can be recovered if the number of parity jumps is faithfully measured. This requires parity measurements to be performed frequently relative to the single photon loss rate. In order to measure the parity of the cavity, we dispersively couple the cavity to an ancilla transmon ($\omega_q / 2\pi$= 6.5 GHz, $T_1^{eg}$ = 26 μs, $T_2^{eg}$ = 12 μs), which is measured using a standard readout chain (See supplementary material 1). When we consider the first three levels of the ancilla ($|g\rangle, |e\rangle$, and $|f\rangle$) this dispersive interaction can be represented as (setting $\hbar = 1$)

$$\hat{H}_{\text{int}} = \chi_e \hat{a}^\dagger \hat{a} |e\rangle\langle e| + \chi_f \hat{a}^\dagger \hat{a} |f\rangle\langle f|, \quad (1)$$

where $\hat{a}$ is the cavity photon annihilation operator, and $\chi_e$, $\chi_f$ are the cavity frequency shifts for the respective ancilla states ($\chi_g$ = 0 in this frame of reference, and in the absence of driving $\chi_e / 2\pi$ = 93 kHz, $\chi_f / 2\pi$ = 236 kHz). Evolution under this interaction for a time $\pi / \chi_e$ = 5.4 μs maps the parity of the cavity onto the phase of a superposition $|g\rangle + |e\rangle$ in the ancilla. Performing Ramsey interferometry on the ancilla to determine this phase yields an effective QND measurement of the

parity (*15, 16*). This parity measurement protocol was previously used to demonstrate error correction at the break-even point (*17*), where the error-corrected lifetime equals that of the best element of the system.

The main limitation of error-correction based on the scheme described above is logical errors induced by spontaneous relaxation of the ancilla during the parity mapping (*17*). This can be seen by considering a jump from $|e\rangle$ to $|g\rangle$ during the $\pi/\chi_e$ interaction time (Fig. 1A). While such a jump prevents one from correctly determining the photon number parity, it also has the more harmful effect of completely dephasing the cavity. Since the jump time is nearly uniformly distributed between 0 and $\pi/\chi_e$, the cavity acquires a phase space rotation uniformly distributed between 0 and $\pi$. This imposes an uncorrectable error with a probability proportional to the number of parity measurements performed. This cost forces the designer of an error correction protocol to measure the error syndrome less frequently than would otherwise be desirable, and consequently reduces the potential achievable lifetime gain. More generally, the non-fault-tolerance of the traditional protocol arises because ancilla relaxation errors do not commute with the interaction Hamiltonian. In particular, the commutator of the interaction Hamiltonian with the associated collapse operator is $\left[\hat{H}_{\text{int}}, |g\rangle\langle e|\right] = -\chi_{eg}\hat{a}^{\dagger}\hat{a}|g\rangle\langle e|$ (where $\chi_{ij} \equiv \chi_i - \chi_j$, for $i, j \in \{g, e, f\}$), which generates a nontrivial operation on the logical subspace, and is therefore an uncorrectable error. In contrast, pure dephasing of the ancilla, which occurs at a comparable rate, does not result in unwanted cavity decoherence because the collapse operator ($|e\rangle\langle e|$) commutes with the interaction. Therefore, the end result of an ancilla dephasing event during the interaction is equivalent to an ancilla dephasing event after the interaction, which clearly does not affect the logical qubit. The parity measurement is therefore "transparent" with respect to ancilla dephasing (*10*).

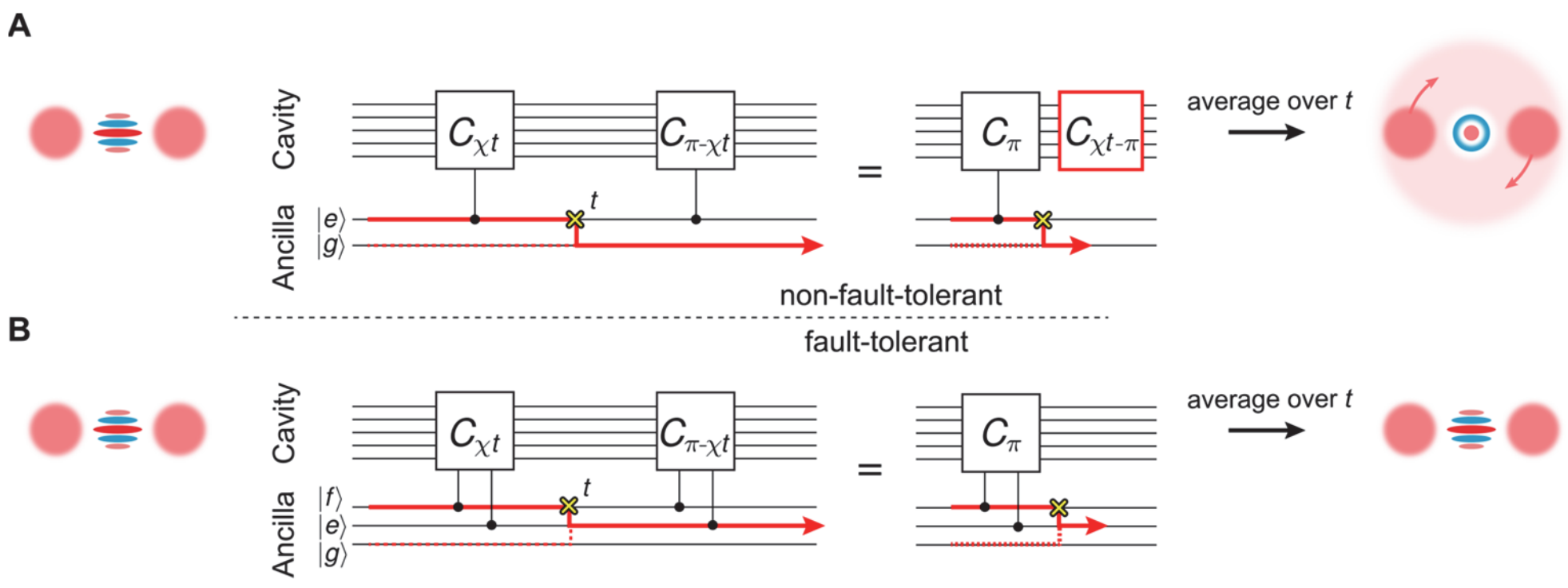

Fig. 1: **Schematic circuit diagram of a FT parity measurement**. Circuit schematic showing the effect of ancilla energy relaxation on a Schrödinger cat state (depicted by its Wigner tomogram, left) during a parity map in both the traditional (A) and FT (B) schemes. In these circuit diagrams the lines within a bundle represent the individual states of the associated mode. $C_\theta = e^{i\theta \hat{a}^\dagger \hat{a}}$ represents a cavity phase shift of angle $\theta$ conditional on the state of the ancilla. **(A)** In the non-FT implementation, an error occurring at time $t \in (0, \pi/\chi)$ results in a cavity phase shift of $\chi t$. This completely dephases the cavity state when averaged over *t*. **(B)** In the FT implementation, an error occurring at time *t* is equivalent to the same error occurring at the end of the parity map, since the error commutes with the interaction.

We extend this error-transparency to include relaxation by introducing a third level to the ancilla Hilbert space (Fig. 1B). This provides us with an additional degree of freedom, allowing us to maintain the system-ancilla interaction rate, while zeroing the rate of first-order error propagation. If we change our initial ancilla encoding to a superposition of $|g\rangle$ and $|f\rangle$ (instead of $|g\rangle$ and $|e\rangle$), the dominant error becomes relaxation from $|f\rangle$ to $|e\rangle$ (selection rules forbid direct $|f\rangle$ to $|g\rangle$ transitions). The commutator of this error ($|e\rangle\langle f|$) with the interaction Hamiltonian is $-\chi_{fe}|e\rangle\langle f|\hat{a}^\dagger\hat{a}$. Because the measurement rate (which scales with $\chi_{fg}$) is independent of the dephasing rate (which scales with $\chi_{fe}$), it becomes feasible to maintain the measurement while removing relaxation-induced dephasing by choosing $\chi_{fg}$ large, and $\chi_{fe} = 0$. The desired FT interaction Hamiltonian is therefore

$$\hat{H}_{\text{int}}^{\text{FT}} = \chi_f \hat{a}^\dagger \hat{a} \left( |e\rangle\langle e| + |f\rangle\langle f| \right), \qquad (2)$$

which clearly commutes with ancilla relaxation from $|f\rangle$ as well as dephasing events.

In our solution, $\chi_{fg}$ is fixed by our sample geometry, and we achieve $\chi_{fe} = 0$ by tuning $\chi_e$ *in situ*. Our tuning is implemented using a sideband tone at a detuning $\Delta$ from the resonant frequency $\omega_{\text{res}} = \omega_{he} - \omega_{\text{c}} = 2\pi \times 8\,\text{GHz}$. This results in a driven sideband term $\hat{H}_{\text{d}} = \frac{\Omega}{2} \hat{a}^\dagger |e\rangle\langle h| e^{i\Delta t} + \text{h.c.}$, which couples the levels $|e, n\rangle$ and $|h, n-1\rangle$ (*18, 19*), with $n$ the number of cavity photons, and $|h\rangle$ the third excited ancilla state (Fig. 2A). For the drive amplitude used throughout this experiment, the single-photon Rabi oscillation rate is $\Omega = 2\pi \times 1.7$ MHz when $\Delta = 0$ (See supplementary material 2). When sufficiently detuned ($\Delta \gg \Omega$) we can approximate this time-dependent Hamiltonian with the time-independent effective interaction:

$$\hat{H}_{\text{eff}} = \frac{\Omega^2}{4\Delta} \left[ \hat{a}^\dagger |e\rangle\langle h|, \hat{a} |h\rangle\langle e| \right] = \chi_e^{\text{ind}} \left[ \left( |e\rangle\langle e| - |h\rangle\langle h| \right) \hat{a}^\dagger \hat{a} - |h\rangle\langle h| \right] \qquad (3)$$

to first order (See supplementary material 3), where $\chi_e^{\text{ind}} = \Omega^2 / 4\Delta$. This Hamiltonian has exactly the form of a dispersive interaction Hamiltonian, conditioned on the ancilla being in $|e\rangle$ or $|h\rangle$. By choosing the detuning, one can engineer an induced $\chi_e^{\text{ind}}$ with either positive or negative sign. Therefore, we consider the total interaction Hamiltonian $\hat{H}_{\text{int}} = \hat{H}_{\text{int}}^0 + \hat{H}_{\text{eff}}$, and the associated dispersive interaction rates $\chi_j = \chi_j^0 + \chi_j^{\text{ind}}$, where the zero index refers to the undriven case. This allows for the total cancellation of either $\chi_{eg}^0$ (at $\Delta = 2\pi\times 9.3$ MHz, Fig. 2B) or $\chi_{fe}^0$ (at $\Delta = 2\pi\times -6.4$ MHz, Fig. 2C), leaving only the higher-order nonlinear dispersive shift of order $\Omega^4/\Delta^3 \ll \chi_e^0, \chi_f^0$ (See supplementary material 3).

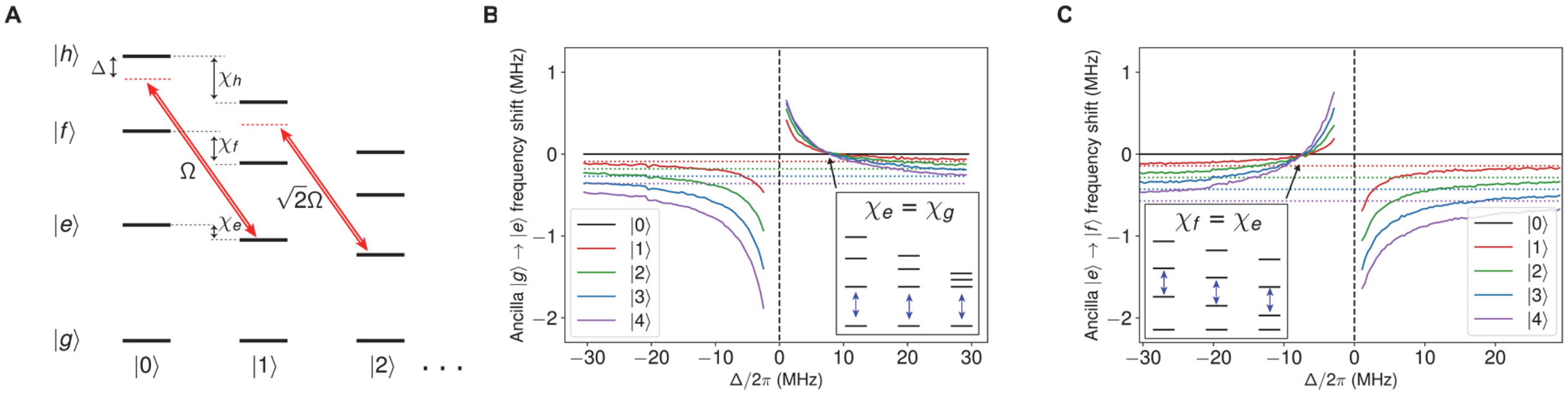

**Fig. 2. Cancelling the dispersive interaction with a sideband drive.** (**A**) Cavity-ancilla level diagram. An applied microwave tone (double red arrows) drives the $|e,n\rangle$, $|h,n-1\rangle$ transition frequency with Rabi rate $\sqrt{n}\Omega$ and detuning $\Delta$. The resulting Stark shift changes the effective $\chi_e$ by an amount $\Omega^2/4\Delta$. (**B**), (**C**) In B (C), spectroscopy of the $|g\rangle$ to $|e\rangle$ ($|e\rangle$ to $|f\rangle$) transition is performed with a varying number of photons in the cavity. $\chi_{eg}$ ($\chi_{fe}$), as well as higher order nonlinear dispersive shifts can be extracted from the spread in transition frequencies with respect to photon number (See supplementary material 3). The indicated crossing points show where $\chi_{eg}$ ($\chi_{fe}$) is approximately zero, as emphasized by the blue arrows in the insets depicting the effective driven level diagram. The dotted lines refer to the transition frequencies when no sideband drive is applied.

We first demonstrate the potential of this approach by using the sideband drive to decouple the cavity from ancilla transitions between $|g\rangle$ and $|e\rangle$ during idle times (Fig. 3A). We achieve this by choosing a detuning such that $\chi_e^{\text{ind}} = -\chi_{eg}^0$, yielding $\chi_{eg} = 0$. This choice of detuning prevents thermal ancilla excitations from $|g\rangle$ to $|e\rangle$ (which occur on average once every 1 ms) from dephasing the cavity, resulting in a dramatic increase in the coherence time of a cavity-encoded qubit (Fig. 3B). If we prepare an initial state $(|0\rangle+|1\rangle)|g\rangle$ and turn on the detuned sideband drive, the coherence increases from $T_2^c \sim 0.7$ ms to 1.9 ms, close to the limit of $2T_1^c \sim$ 2.2 ms. This is possible because the cavity lifetime is almost completely unaffected by the drive when the ancilla is in the ground state. The residual dephasing time of $T_\phi^c$ =14 ± 1 ms can mostly be explained by second-order excitations from $|e\rangle$ to $|f\rangle$. This demonstration not only showcases the effectiveness of the drive in cancelling the system-ancilla interaction, but also shows that the addition of the drive does not produce unwanted cavity decoherence at an appreciable level.

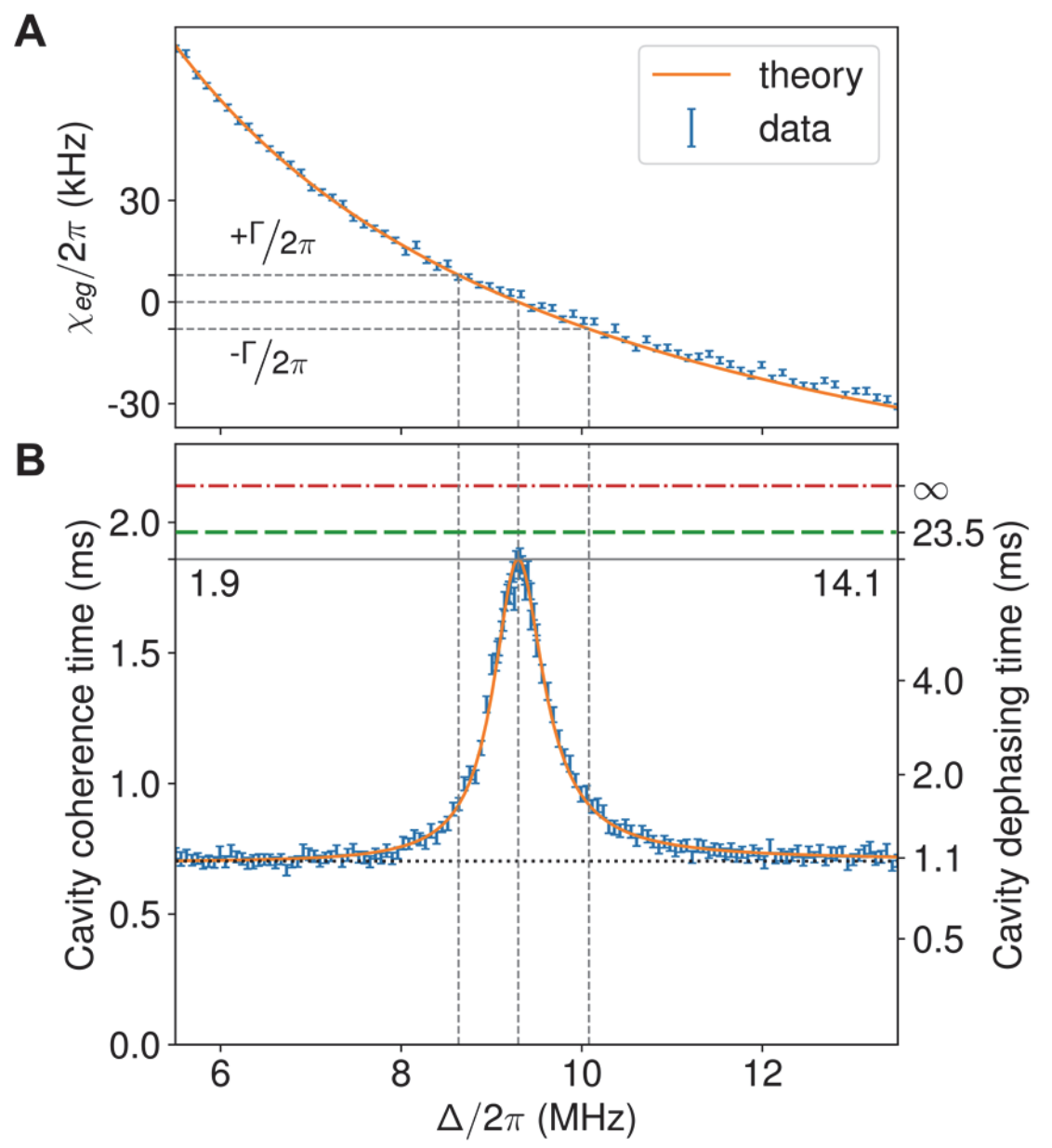

Fig. 3: **Improving the cavity coherence time by decoupling the cavity from thermal ancilla excitations.** While a bare cavity is nearly completely limited by single photon loss, a cavity dispersively coupled to an ancilla experiences dephasing as a result of spontaneous ancilla excitation. (**A**) The measured dispersive interaction (blue markers) varies as a function of sideband drive detuning from resonance Δ as $\chi_{eg} = \chi_{eg}^0 + \Omega^2 / 4\Delta$ (solid orange line). (**B**) Cavity coherence times as a function of the sideband drive frequency obtained from cavity Ramsey experiments. In the absence of quantum error correction, the cavity coherence time is ultimately limited to $2T_1^c \sim 2.2$ ms (red dot-dashed line). Without sideband drive, thermal ancilla excitations limit the cavity coherence to about 700 μs (dotted black line). This dephasing source is almost entirely removed for $\chi_{eg} = 0$. The remaining dephasing time (i.e. excluding photon decay) is $T_\phi^c$ =14 ± 1 ms (solid grey line),

close to the 23.5 ms limit set by second-order thermal excitation from $|e\rangle$ to $|f\rangle$ (dashed green line). The analytical behavior of the cavity coherence (orange line, see supplementary material 4) closely matches the observed values. Protection against thermal ancilla excitations starts occurring when $|\chi_{eg}| < \Gamma / 2\pi$ (dashed grey lines), where $\Gamma = 1/T_1^{eg}$ is the qubit $|e\rangle$ to $|g\rangle$ decay rate.

Next, we construct the FT parity measurement protocol by choosing the appropriate detuning Δ, such that $\chi_e^{\mathrm{ind}} = +\chi_{fe}^0$ and therefore $\chi_{fe} = 0$. In this case we realize the Hamiltonian of Eq. 2, for which ancilla relaxation from $|f\rangle$ to $|e\rangle$ does not change the evolution of the cavity. In order to qualitatively demonstrate the resulting fault tolerance, we follow the protocol in Fig. 4A with $N$ = 1. In this experiment, we first prepare an even Schrödinger cat state with mean photon number two in the cavity (See supplementary material 5). We then map the photon-number parity onto the ancilla in three different ways (Fig. 4B-D), as outlined below. We measure the ancilla to determine the outcome of the parity measurement, and reset it to the ground state. Finally, we perform Wigner tomography on the cavity in order to determine the fidelity of the final cavity state conditioned on the outcome of the first parity measurement. In order to focus on ancilla-induced errors, we filter out instances in which a photon loss event occurred (See supplementary material 6).

We demonstrate the advantage of the FT protocol ($\Pi_{\mathrm{FT}}$) by comparing it with two alternative protocols: the traditional parity measurement ($\Pi_{ge}$), which uses a $|g\rangle + |e\rangle$ encoding in the ancilla, and $\Pi_{gf}$, which uses a $|g\rangle + |f\rangle$ encoding, but without applying the sideband drive that zeroes $\chi_{fe}$. All three protocols have similar parity assignment fidelities of 83%, 86.5% and 82% for $\Pi_{ge}, \Pi_{gf}$ and $\Pi_{\mathrm{FT}}$, respectively. In the absence of photon loss, the outcome of the parity measurement indicates specific ancilla events during the parity mapping (See supplementary material 7). In the traditional $\Pi_{ge}$ parity mapping protocol, the outcome is either $|g\rangle$ or $|e\rangle$ (Fig. 4B). No-error events result in $|g\rangle$, whereas dephasing events lead the ancilla to end up in $|e\rangle$. Relaxation errors cannot be singled out, as they result in a detection of $|g\rangle$ or $|e\rangle$ with equal probability. Relaxation errors therefore manifest as a lowered fidelity of the cavity state for both outcomes, a direct consequence of non-fault-tolerance. We next perform the $\Pi_{gf}$ protocol, without applying the sideband drive (Fig. 4C). To initialize the ancilla in a $|g\rangle+|f\rangle$ superposition, we use a *g-e* $\pi/2$-pulse followed by a *e-f* $\pi$-pulse. We then allow the system to evolve under the interaction Hamiltonian for a time $\pi/\chi^0_{fg}\sim 2$ μs so that the cavity phase space acquires a conditional $\pi$ rotation. After applying the reverse of the ancilla preparation sequence, the ancilla is in state $|g\rangle$ if no ancilla error has occurred. If a dephasing error occurs, the ancilla ends up in $|e\rangle$. In contrast to the $\Pi_{ge}$ protocol, we can now distinguish relaxation events, for which the ancilla ends up in $|f\rangle$. It is now evident that dephasing events do not affect the cavity state (Fig. 4C), whereas a relaxation event, which does not commute with the interaction, dephases the cavity state.

Finally, we perform the FT parity mapping $\Pi_{\mathrm{FT}}$ (Fig. 4D). In addition to the sequence of the $\Pi_{gf}$ protocol, we now also apply the sideband drive so that $\chi_{fe}=0$ in the time period between the two *e-f* $\pi$-pulses. In this case, we see that the cavity coherence is maintained even in the case of ancilla relaxation. The modest increase in the prevalence of dephasing events is a result of a slightly degraded ancilla dephasing rate in the presence of the strong drive.

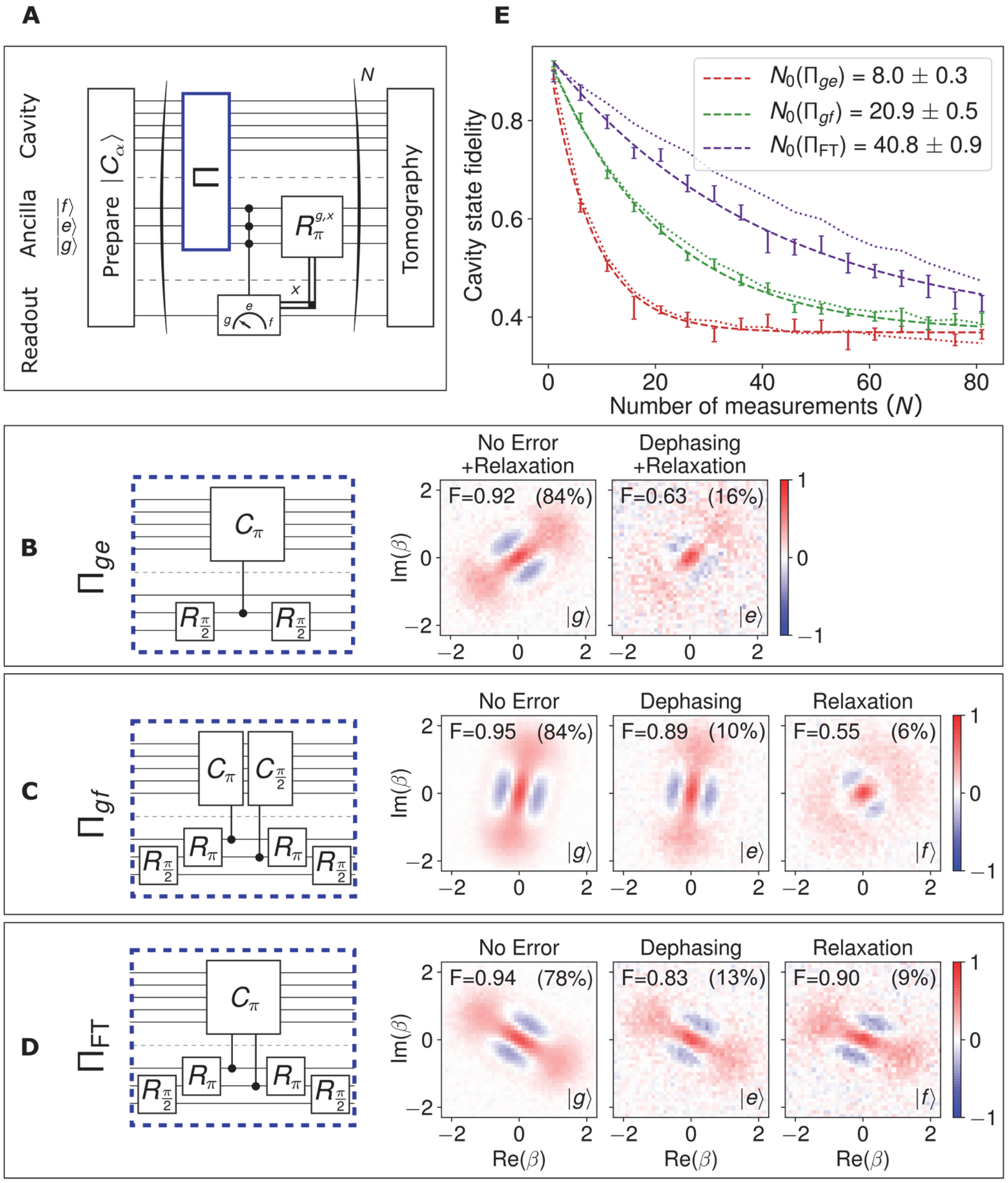

**Fig. 4. Demonstration of a fault-tolerant parity measurement. (A).** Circuit protocol characterizing the parity syndrome measurement and tomography (See supplementary material 5-6) of the resulting cavity state. We start by initializing the cavity in $|C_\alpha\rangle$. After every parity map $\Pi$ (indicated in blue), we perform a three-outcome ancilla readout, and reset the ancilla using $\pi$-pulses ($R_\pi$). The parity measurements are implemented in three different ways as shown in (B, C, D). In order to focus on ancilla errors, the tomography includes parity measurements used to filter out photon loss. **(B, C, D)**. Wigner tomography of the cavity state conditioned on the outcome of a single parity measurement ($N$=1). The outcome (shown in the bottom right of each Wigner plot) informs us about ancilla behavior during the parity mapping (top, see supplementary material). The prevalence of this outcome is indicated in the top right. For each Wigner tomogram, a state fidelity F (shown in the top left) is given, each with statistical error smaller than 0.01. The fidelity of the initial cat state is ~0.95 due to imperfections in state preparation and tomography. For $\Pi_{ge}$ (B) and $\Pi_{gf}$ (C), ancilla relaxation results in a dephased cavity state, whereas for $\Pi_{\mathrm{FT}}$ (D) the logical qubit is preserved. **(E).** Fidelity vs. number of measurements (*N*) for the three types of parity measurement. The dotted lines are simulated fidelities extracted from Monte-Carlo trajectories (See supplementary material), and the dashed lines are exponential fits to the data $F(N) = Ae^{-N/N_0} + c$ with $A \sim 0.56$ and $c \sim 0.37$ for all curves.

In an error-correction setting, the parity of the logical qubit must be repeatedly measured. In order to demonstrate the advantage supplied by the FT parity measurement in this context, we use the protocol indicated in Fig. 4A, and extract the final state fidelity as a function of the number of measurements (*N*). With an exponential fit, we can assign a characteristic number of measurements ($N_0$) in which the cavity fidelity decays. At this point, we can quantify the improvement offered by the FT protocol. We see that $N_0(\Pi_{gf}) / N_0(\Pi_{ge}) = 2.6 \pm 0.2$, showing that even without sideband drive, the $\Pi_{gf}$ protocol offers some advantages compared to $\Pi_{ge}$. The first reason is that the probability of relaxation is lower for $\Pi_{gf}$, since the relaxation time of $|f\rangle$ (24 μs) is nearly that of $|e\rangle$ (26 μs), while the parity measurement time of $\Pi_{gf}$ is less than half that of $\Pi_{ge}$. The second reason is that the cavity is less dephased given that an ancilla relaxation event occurred, since the cavity angle is distributed between 0 and $\pi\chi_{fe}^{0} / \chi_{fg}^{0} = 0.6\pi$ (As evident from the residual coherence after a relaxation event in Fig. 4C). The FT implementation improves on $\Pi_{gf}$ by a factor of 2.0 ± 0.1, resulting in a total fault-tolerance gain of $N_0(\Pi_{\mathrm{FT}}) / N_0(\Pi_{ge}) =$ 5.1 ± 0.3. We can compare the observed cavity dephasing rates with predictions for residual uncorrected errors, the largest of which are thermal excitation during the parity map and decay during readout (See supplementary material 9). Monte-Carlo simulations (See supplementary material 8) of how the cavity phase distribution is affected by these factors produce fidelity decay curves which are in good agreement with the observed results.

It is worth emphasizing the distinction between the FT implementation of an operation, as demonstrated here, and FT quantum computing architectures. While the former is assessed in terms of reduction of error propagation, the latter is commonly interpreted as the presence of an error threshold, below which the error rate of a system scales favorably with system size. Any FT architecture must contain FT syndrome measurements, and therefore our results are a necessary step toward realizing such a system. While our implementation differs from those proposed in traditional schemes, it offers several advantages. Because errors are prevented from propagating to the system, it does not require a subsequent round of error correction to recover from ancilla-induced errors. In addition, the syndrome measurement is also transparent to photon loss, and is therefore fully compatible with the cat-code error correction scheme. The presented scheme is readily extendable to higher orders of FT protection. For instance, by using four instead of three ancilla levels, we can protect against relaxation errors up to second order, or alternatively against both relaxation and thermal excitations to first order. However, moving to higher orders of protection is challenging, since additional drives may ultimately degrade the system coherence. In an alternative proposal, using an approach termed degeneracy-preserving measurement (*20*), the interaction could be protected from ancilla errors of any order, providing the strongest form of fault-tolerance.

To summarize, we demonstrated a parity check syndrome measurement of a logical qubit that is protected against all first-order ancilla errors, yielding a fault-tolerance gain of five compared to the non-FT measurement. We hope to adapt the techniques used here to produce both improved syndrome measurements, with higher-order levels of fault-tolerance, as well as additional FT computing operations. We envision the hardware-efficient approach presented in this work to become an important tool in the development of a complete FT computing architecture.

**Acknowledgments:** We thank Isaac Chuang and Loïc Herviou for helpful discussions, Christopher S. Wang and Shantanu Mundhada for reviewing the manuscript, and N. Ofek for providing the logic for the field programmable gate array (FPGA) used for the control of this experiment. This research was supported by the U.S. Army Research Office (W911NF-14-1-0011). P.R. was supported by the U.S. Air Force Office of Scientific Research (FA9550-15-1-0015); L.J. by the Alfred P. Sloan Foundation and the Packard Foundation. Facilities use was supported by the Yale Institute for Nanoscience and Quantum Engineering (YINQE), the Yale SEAS cleanroom, and the National Science Foundation (MRSECDMR-1119826).

## Supplementary Materials

### 1. Experimental Setup

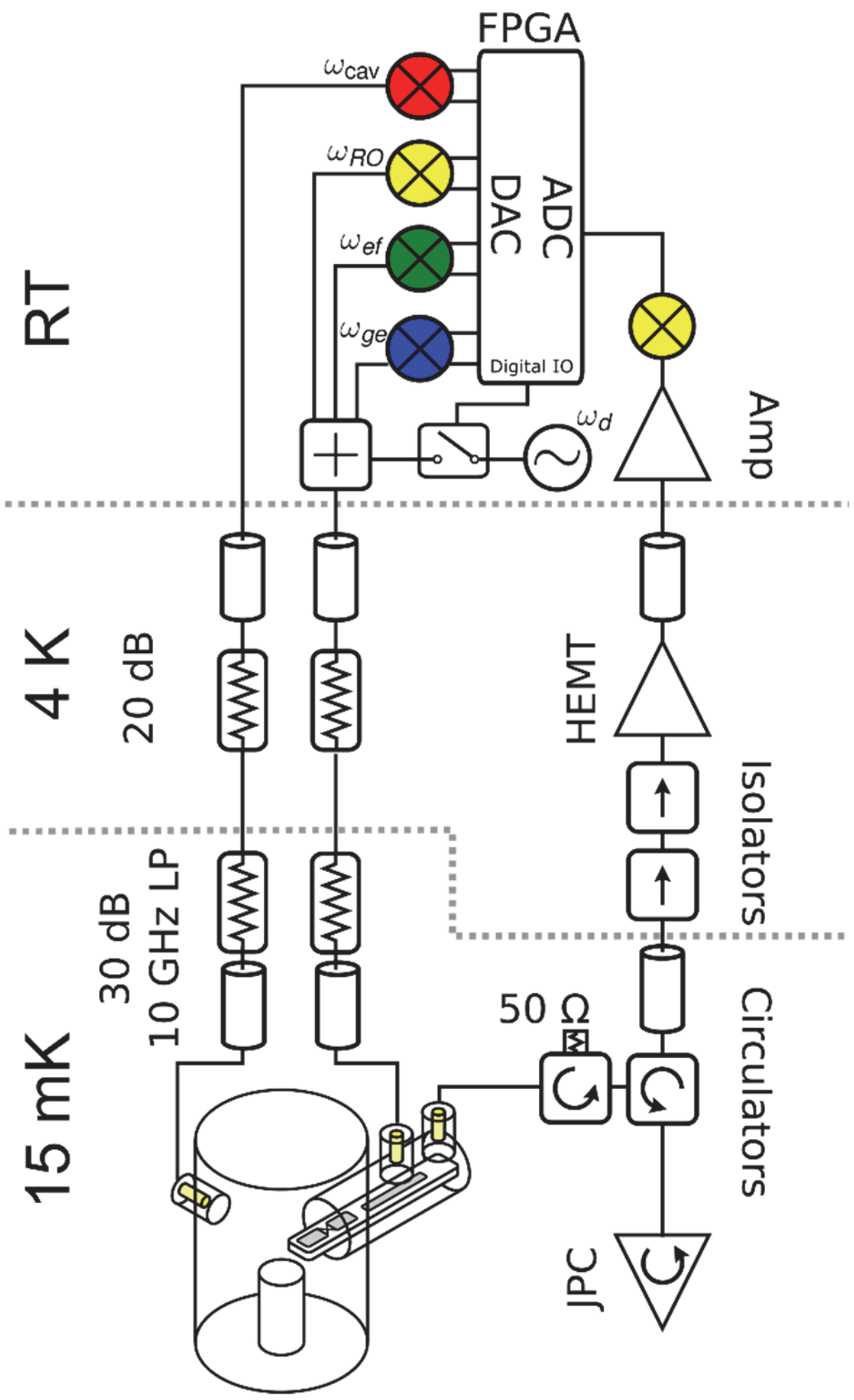

**Fig. S1:** An FPGA controller (2x Innovative Integration X6-1000M in VPXI-ePC chassis) generates 4 pairs of I/Q waveforms using 500 Msample/s digital to analog converters (DAC). Each pair is upconverted using an I/Q mixer (Marki IQ-0307-LXP or IQ-0618-LXP depending on the frequency). The two yellow mixers share the same local oscillator. To prevent problems due to mixer leakage, each local oscillator is set 50 MHz above the desired frequency and single-sideband modulation is used. Additionally, a separate RF generator (Agilent MXG N5183A) is gated by a digital pulse provided by the FPGA to enable or disable the detuned sideband drive. Proper attenuation at each temperature stage is crucial to thermalize the blackbody radiation from the 50Ω environment. Additional low-pass filters (K&L250-10000 and home-built eccosorb) protect the sample from spurious high-frequency components. The output chain consists of a Josephson Parametric Converter (JPC), which reflects the input signal with ~20 dB of gain (bandwidth ~3 MHz). The circulators (Pamtech XTE0812KC) prevent the amplified signal from going back to the sample and direct it through 2 isolators (Pamtech CWJ0312KI) to a HEMT-amplifier (Low Noise Factory LNF-LNR1 12A). Finally, an image reject mixer (Marki SSB-0618) converts the RF signal back to the intermediate frequency (50 MHz). The FPGA samples the signal using a 1 Gsample/s analog to digital converter (ADC), demodulates and integrates to indicate whether the transmon was in $|g\rangle$, $|e\rangle$ or $|f\rangle$.

## 2. Deriving the Sideband Hamiltonian

The full Hamiltonian for a single Josephson junction (Josephson energy $E_{\mathrm{J}}$) coupled to two modes ($\hat{a}$ for the cavity and $\hat{q}$ for the transmon) with a drive at frequency $\omega_{\mathrm{d}}$ can be written as

$$H^{(0)} = \omega_{\mathrm{c}}^{(0)}\hat{a}^{\dagger}\hat{a} + \omega_{\mathrm{q}}^{(0)}\hat{q}^{\dagger}\hat{q} + E_{\mathrm{J}}\cos\left(\phi_{\mathrm{c}}(\hat{a}+\hat{a}^{\dagger}) + \phi_{\mathrm{q}}(\hat{q}+\hat{q}^{\dagger})\right) + \varepsilon\cos(\omega_{\mathrm{d}}t)(\hat{q}+\hat{q}^{\dagger}) .$$

Second-order terms in the cosine provide a hybridization of the two modes, which can be accounted for by using the "dressed" modes. Sixth- and higher-order terms of the cosine can be neglected, leaving

$$H^{(0)} \approx \omega_{\mathrm{c}}\hat{a}^{\dagger}\hat{a} + \omega_{\mathrm{q}}\hat{q}^{\dagger}\hat{q} + \frac{E_J}{24}\left(\phi_{\mathrm{c}}(\hat{a}+\hat{a}^{\dagger}) + \phi_{\mathrm{q}}(\hat{q}+\hat{q}^{\dagger})\right)^4 + \varepsilon\cos(\omega_{\mathrm{d}}t)(\hat{q}+\hat{q}^{\dagger}) .$$

The time dependence of the drive can be eliminated by going into a frame rotating at $\omega_{\mathrm{d}}$ and dropping counter-rotating terms,

$$\begin{aligned} H^{(1)} &= e^{i\omega_{\mathrm{d}}t\hat{q}^{\dagger}\hat{q}}H^{(0)}e^{-i\omega_{\mathrm{d}}t\hat{q}^{\dagger}\hat{q}} - \omega_{\mathrm{d}}\hat{q}^{\dagger}\hat{q} \\ &= \omega_{\mathrm{c}}\hat{a}^{\dagger}\hat{a} + (\omega_{\mathrm{q}}-\omega_{\mathrm{d}})\hat{q}^{\dagger}\hat{q} + \frac{E_J}{24}\left(\phi_{\mathrm{c}}\hat{a} + \phi_{\mathrm{q}}\hat{q}e^{i\omega_{\mathrm{d}}t} + \text{h.c.}\right)^4 + \varepsilon(\hat{q}+\hat{q}^{\dagger}) + O(e^{2i\omega_{\mathrm{d}}t}) \end{aligned}$$

The drive term can be eliminated (assuming $\varepsilon$ quasi-static) by going into a displaced frame with displacement $\xi = \dfrac{-\varepsilon}{\omega_{\mathrm{q}}-\omega_{\mathrm{d}}}$ resulting in the new Hamiltonian,

$$\begin{aligned} H^{(2)} &= e^{i\xi(\hat{q}+\hat{q}^{\dagger})}H^{(1)}e^{-i\xi(\hat{q}+\hat{q}^{\dagger})} \\ &= \omega_{\mathrm{c}}\hat{a}^{\dagger}\hat{a} + (\omega_{\mathrm{q}}-\omega_{\mathrm{d}})\hat{q}^{\dagger}\hat{q} + \frac{E_J}{24}\left(\phi_{\mathrm{c}}\hat{a} + \phi_{\mathrm{q}}(\hat{q}+\xi)e^{i\omega_{\mathrm{d}}t} + \text{h.c.}\right)^4 \end{aligned}$$

Finally, by going into the interaction frame, we are left with just the effective nonlinear terms.

$$H^{(3)} = \frac{E_J}{24}\left(\phi_c \hat{a} e^{i\omega_c t} + \phi_q \hat{q} e^{i\omega_q t} + \phi_q \xi e^{i\omega_d t} + \text{h.c.}\right)^4.$$

Now, assuming no unintentional frequency collisions, after dropping high-frequency oscillating terms all that remains are the diagonal terms (functions of $\hat{a}^\dagger \hat{a}$ and $\hat{q}^\dagger \hat{q}$) and the desired sideband interaction

$$H^{(3)} \approx H_c(\hat{a}^\dagger \hat{a}) + H_q(\hat{q}^\dagger \hat{q}) + H_{qc}(\hat{a}^\dagger \hat{a}, \hat{q}^\dagger \hat{q}) + \frac{\widetilde{\Omega}}{2}(\hat{a}^\dagger \hat{q}\hat{q} e^{i\tilde{\Delta} t} + \text{h.c.}),$$

where the sideband drive strength is $\widetilde{\Omega} = E_J \xi \phi_c \phi_q^3$ and the detuning $\tilde{\Delta} = 2\omega_q - \omega_c - \omega_d$. Since $H_c, H_{qc} \ll \widetilde{\Omega} \ll H_q$, the induced sideband transitions are unselective with respect to the cavity state, but selective with respect to the transmon state. Defining $\Delta \equiv \tilde{\Delta} - \langle h | H_q | h \rangle + \langle e | H_q | e \rangle$, we can make a final simplification:

$$\begin{aligned} H^{(4)} &= e^{iH_q t} H^{(3)} e^{-iH_q t} - H_q \\ &\approx H_c(\hat{a}^\dagger \hat{a}) + H_{qc}(\hat{a}^\dagger \hat{a}, \hat{q}^\dagger \hat{q}) + \frac{\Omega}{2}(\hat{a}^\dagger | e \rangle\langle h | e^{i\Delta t} + \text{h.c.}) \end{aligned}$$

where $\Omega = \sqrt{6}\widetilde{\Omega}$. When switching on the drive at the expected frequency, we indeed observe Rabi oscillations corresponding to the sideband transition $|e, n\rangle \leftrightarrow |h, n-1\rangle$ with sideband Rabi rate $\Omega = 2\pi \times 1.7$ MHz for $n = 1$ (Fig. S2). The fact that this sideband interaction is much larger than $\chi_{fe}^0$ and $\chi_{eg}^0$ is a precondition for its application as a $\chi$-cancelling drive.

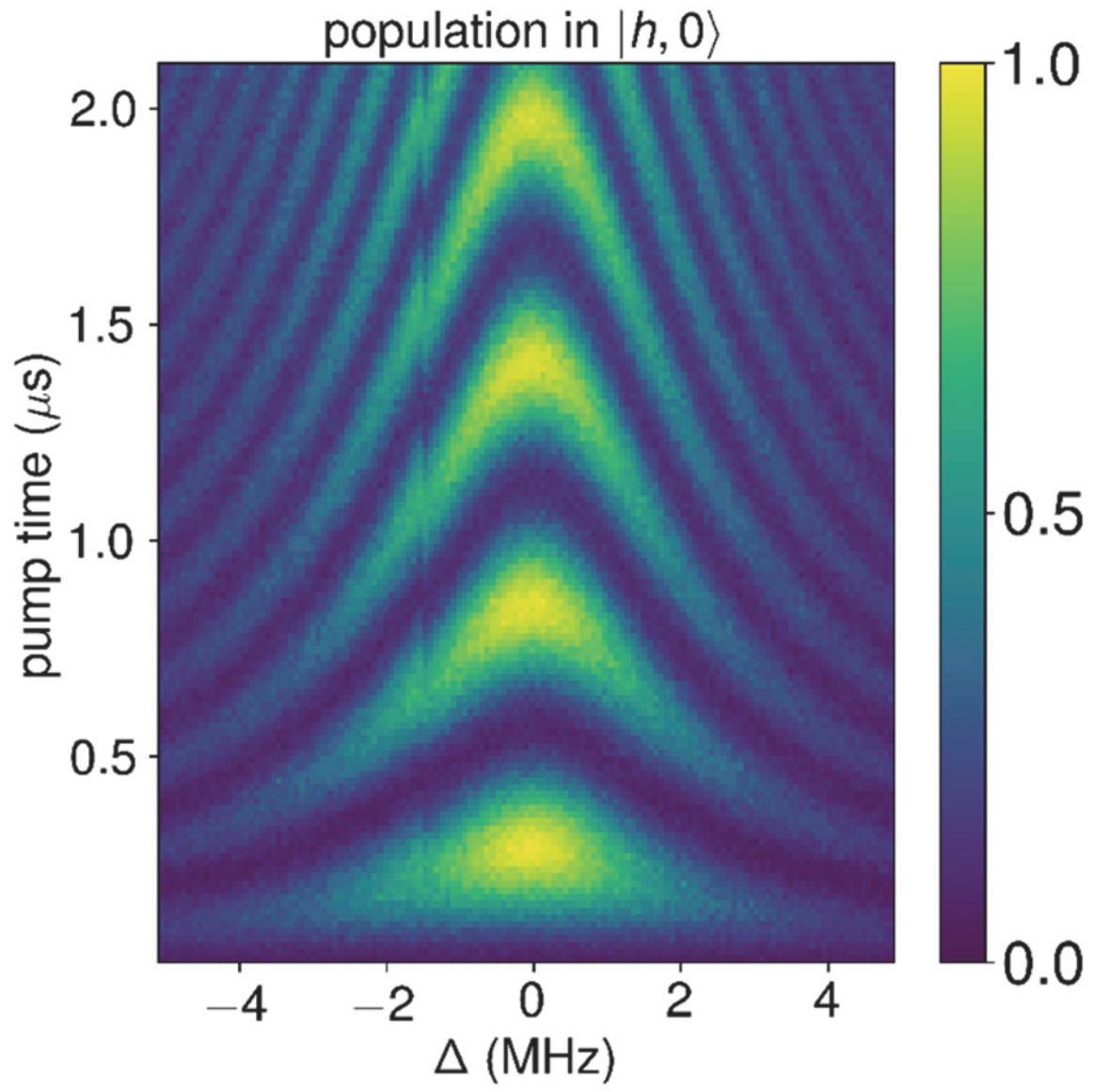

**Fig. S2:** Chevron pattern observed in the population of $|h,0\rangle$ when preparing $|e,1\rangle$ and switching on the sideband drive at a detuning $\Delta$ from resonance for a varying amount of time.

## 3. Deriving the Effective Hamiltonian

Given a time dependent Hamiltonian of the form $H(t) = g(\hat{A}e^{i\Delta t} + \text{h.c.})$, if sufficiently detuned from resonance (*i.e.* $g \ll \Delta$) then it becomes possible to replace the time dependent Hamiltonian with a static, effective Hamiltonian $H_{\text{eff}}^{(1)} = \frac{g^2}{\Delta}[\hat{A}, \hat{A}^\dagger]$. The form of this effective Hamiltonian can be derived either from operator-based Floquet theory (*21*), or using the second-order rotating wave approximation (*22*). In the case of the detuned sideband drive derived in the previous section, $H_{\text{eff}}$ gives a dispersive shift as seen in Eq. 3 of the main text. The next lowest order term which is resonant provides a non-linear dispersive shift (Fig. S3) $H_{\text{eff}}^{(2)} \propto \frac{g^4}{\Delta^3}[[[\hat{A}, \hat{A}^\dagger], \hat{A}], \hat{A}^\dagger] = \chi_e' \hat{a}^\dagger \hat{a}^\dagger \hat{a}\hat{a}(|e\rangle\langle e| - |h\rangle\langle h|)$.

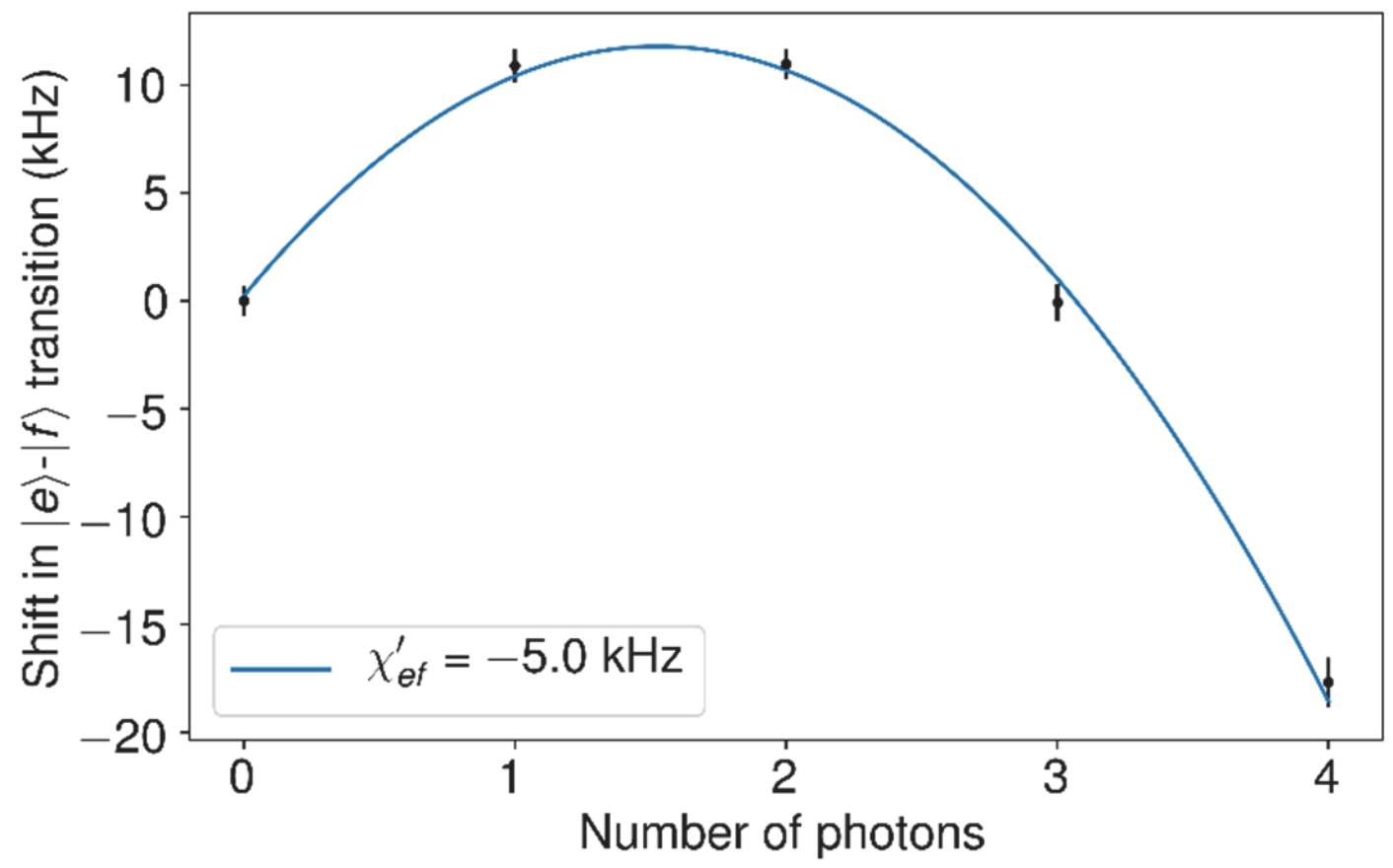

**Fig. S3: Measurement of the non-linear dispersive shift**. With the off-resonant sideband drive applied, the $|e\rangle$-$|f\rangle$ transition becomes nearly independent of photon number. However, a small non-linear term of the form $\chi_e' \hat{a}^\dagger \hat{a}^\dagger \hat{a}\hat{a}\,|e\rangle\langle e|$ remains. The size of this term can be extracted from the quadratic fit (shown in blue) to the measurements (in black) of the $|e\rangle$-$|f\rangle$ transition frequency as a function of photon number.

## 4. Analytic Model for Cavity Dephasing due to Thermal Excitation

In the dispersive interaction regime, the presence of a thermal population $\bar{n}_{\text{th}}^{e}$ in the ancilla leads to frequency fluctuations in the cavity, and hence to a dephasing rate of

$$\frac{1}{T_{\phi,eg}^{\text{c}}}=\frac{\Gamma}{2}\text{Re}\left(\sqrt{\left(1+\frac{i\chi_e}{\Gamma}\right)^2+\frac{4i\chi_e\bar{n}_{\text{th}}^{e}}{\Gamma}}-1\right),$$

where $\Gamma=1/T_1^{eg}$ (*23*). In the limit of large $|\chi_e|/\Gamma$, thermal ancilla excitations last long enough on average to completely dephase the cavity, leading to a $\chi_e$-independent dephasing time of $T_{\phi,eg}^{\text{c}}\left(|\chi_e|\gg\Gamma\right)=T_1^{eg}/\bar{n}_{\text{th}}^{e}$, as evident from the constant background in Fig. 3b of the main text. In the limit of small $|\chi_e|/\Gamma$, the cavity phase undergoes a small random rotation for every ancilla excitation. The spread in this rotation leads to a dephasing time of $T_{\phi,eg}^{\text{c}}\left(|\chi_e|\ll\Gamma\right)=\left(\chi_e^2T_1^{eg}\bar{n}_{\text{th}}^{e}\right)^{-1}$ for $\bar{n}_{\text{th}}^{e}\ll 1$, resulting in a sharply rising peak as we approach $\chi_e=0$ in Fig. 3b. The cavity coherence time $T_2^{\text{c}}(\chi_e)=1/\left[1/2T_1^{\text{c}}+1/T_\phi^{\text{c}}(\chi_e)\right]$ with $1/T_\phi^{\text{c}}(\chi_e)=1/T_{\phi,eg}^{\text{c}}(\chi_e)+1/T_{\phi,\text{res}}^{\text{c}}$ is used for the analytical curve of Fig. 3b, yielding an accurate model for the measured data.

The measured residual dephasing $T_{\phi,\text{res}}^{\text{c}}$ =14 ± 1 ms is partially accounted for by double ancilla excitations from $|g\rangle$ to $|f\rangle$, leading to a predicted residual dephasing time of $T_1^{eg}/2\left(\bar{n}_{\text{th}}^{e}\right)^2$ = 23.5 ms.

## 5. Cat State Preparation Protocol

While there are deterministic protocols available for creating Schrödinger cat states using the dispersive interaction (*12*), for simplicity, in this experiment, we prepare cat states by displacing, measuring parity, and postselecting on even parity. In order to maximize the preparation fidelity, we measure parity four times (using the $\Pi_{gf}$ protocol), resulting in a preparation success rate of 33%, and a final photon number parity of 99%. The remaining probability of odd parity is partly due to the finite parity assignment fidelity, and partly because of the 0.4% probability of a photon jump occurring during each parity measurement.

## 6. Wigner Tomography and State Reconstruction

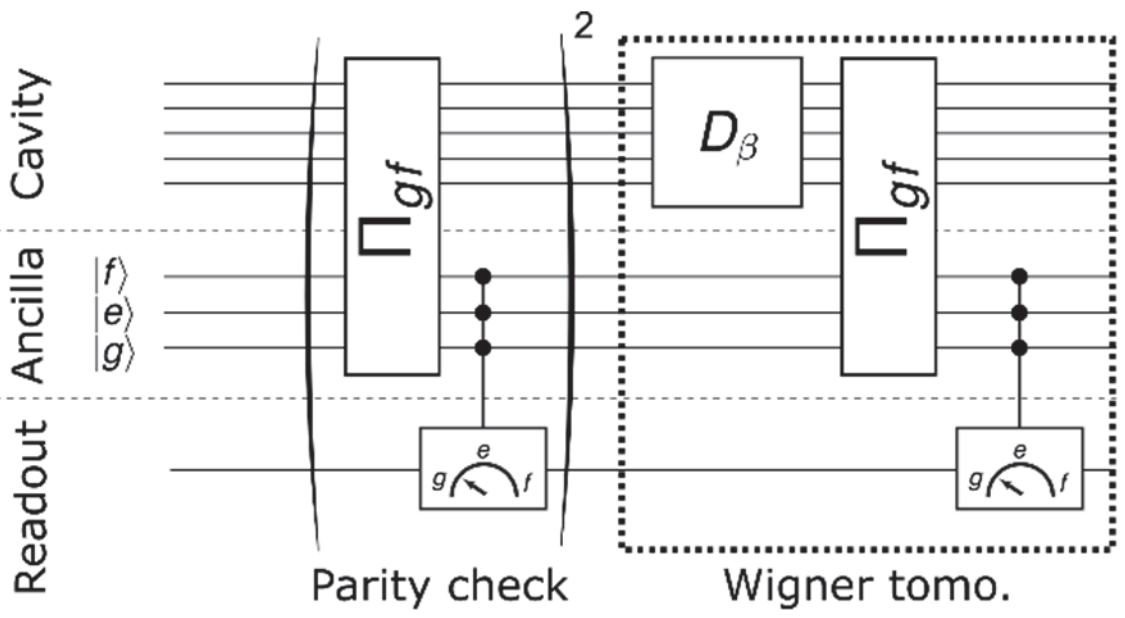

**Fig. S4: Wigner Tomography Circuit.** The circuit consists of parity-mappings ($\Pi_{gf}$) cavity displacements (*D*) and three-outcome ancilla readout. In the single parity measurement data (Fig. 4B-D), two parity checks are used to filter out photon loss, which are in principle correctable in the Schrödinger cat code. In the repeated parity measurement experiment, the trajectory of parity measurement outcomes itself is used to filter out photon loss (see below).

In order to calculate the state fidelity as done in Fig. 4 in the main text, we first perform postselection in order to remove photon loss events, and then do Wigner tomography (Fig. S4). The data from Fig. 4B-D is kept only if two subsequent parity checks indicate even parity. This results in discarding approximately 30% of the data, since the parity measurement assignment fidelity is 85%. In the repeated parity measurement (Fig. 4E, main text), the post-selection is performed in a different way. Because the parity can change multiple times during the experiment, the final state parity is not a good indicator of whether photons were lost. Therefore, we instead compute the probability of obtaining the full record of parity measurement results, for each shot, assuming no photons were lost. If this probability was sufficiently small ($p < 20\%$) then the data was discarded. This results in between 10% and 50% of the data being discarded as $N$ varies from 1 to 80.

We retrieve the maximum likelihood density matrix ( $\rho$ ) after normalizing the Wigner function by the measured parity of the vacuum state ( $\langle 0|\Pi|0\rangle \approx 0.735$ ) to remove the effect of measurement infidelity. The resulting state is constrained to be positive semidefinite and have trace one. Finally, we apply a phase space rotation which maximizes the fidelity to the fiducial state $|C_\alpha^+\rangle \propto |\alpha\rangle + |-\alpha\rangle$ . The fidelity is calculated as $|\langle C_\alpha^+|\rho|C_\alpha^+\rangle|$ .

## 7. Step-by-step Analysis of Parity Measurements

The outcomes of parity measurements inform us not only about the occurrence of a photon loss, but may also be correlated with specific ancilla errors, as specified in Fig. 4 of the main manuscript. In Tables S1 and S2, a detailed analysis of the ancilla state evolution is presented for various types of events. The cavity has been assumed to be in an even parity state and is omitted in this description for clarity.

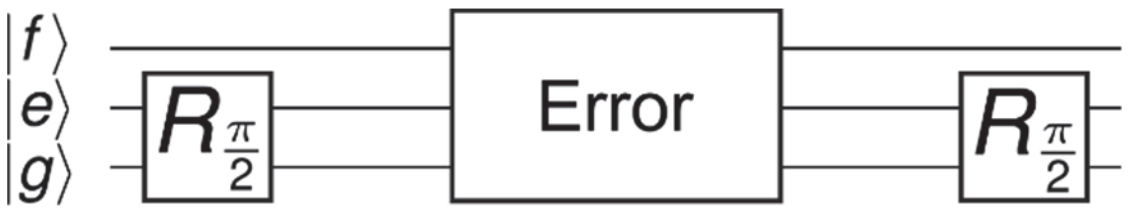

| | Initial state | $R^{ge}_{\pi/2}$ | Error | $R^{ge}_{\pi/2}$ (Final state) |
|---|---|---|---|---|
| **No error** | $\lvert g\rangle$ | $\lvert g\rangle+\lvert e\rangle$ | $\lvert g\rangle+\lvert e\rangle$ | $\lvert g\rangle$ |
| **Dephasing** | $\lvert g\rangle$ | $\lvert g\rangle+\lvert e\rangle$ | $\lvert g\rangle-\lvert e\rangle$ | $\lvert e\rangle$ |
| **Relaxation** | $\lvert g\rangle$ | $\lvert g\rangle+\lvert e\rangle$ | $\lvert g\rangle$ | $\lvert g\rangle+\lvert e\rangle$ |

**Table S1: Detailed analysis of the traditional parity measurement ($\Pi_{ge}$).** The no-error case is distinguishable from a dephasing error. Relaxation errors, on the other hand, cannot be singled out, and lead to $\lvert g\rangle$ and $\lvert e\rangle$ measurements with equal probabilities. Photon loss, if not filtered out, is indistinguishable from a dephasing error.

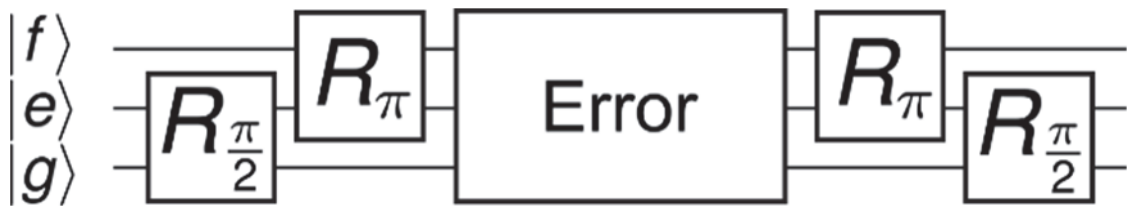

| | Initial state | $R^{ge}_{\pi/2}$ | $R^{ef}_{\pi}$ | Error | $R^{ef}_{\pi}$ | $R^{ge}_{\pi/2}$ (Final state) |
|---|---|---|---|---|---|---|
| **No error** | $\|g\rangle$ | $\|g\rangle+\|e\rangle$ | $\|g\rangle+\|f\rangle$ | $\|g\rangle+\|f\rangle$ | $\|g\rangle-\|e\rangle$ | $\|g\rangle$ |
| **Dephasing** | $\|g\rangle$ | $\|g\rangle+\|e\rangle$ | $\|g\rangle+\|f\rangle$ | $\|g\rangle-\|f\rangle$ | $\|g\rangle+\|e\rangle$ | $\|e\rangle$ |
| **Relaxation** | $\|g\rangle$ | $\|g\rangle+\|e\rangle$ | $\|g\rangle+\|f\rangle$ | $\|e\rangle$ | $\|f\rangle$ | $\|f\rangle$ |

**Table S2: Detailed analysis of the three-level parity measurements (** $\Pi_{gf}$ **and** $\Pi_{\mathrm{FT}}$ **).** Dephasing, relaxation, and no error can be distinguished as they lead to different final ancilla states. Photon loss, if not filtered out, is again indistinguishable from a dephasing error.

## 8. Monte-Carlo Simulation

In order to simulate the fidelity of the cavity state after a sequence of parity measurements, as done in Fig. 4E in the main text, we use a model that takes into account the errors listed in Table S3. Each of the $k$ errors has a probability of occurrence ($p_1,\ldots,p_k$), a change in cavity frequency ($\delta\chi^{(1)},\ldots,\delta\chi^{(k)}$), and a range of times for which this shift is active ($[t_0^{(1)},t_1^{(1)}],\ldots,[t_0^{(k)},t_1^{(k)}]$). We simulate the cavity's trajectory over $N$ parity measurements by sampling the number of each event from the multinomial distribution $n_1,\ldots,n_k \sim \text{Multinomial}(N, p_1,\ldots,p_k)$. Then, for each event, we sample the change in cavity phase from a uniform distribution associated with that event $\theta_{i,j} \sim \text{Unif}(\delta\chi^{(i)}t_0^{(i)}, \delta\chi^{(i)}t_1^{(i)})$. Finally, we sum these phases $\theta = \sum_{i=1}^{k}\sum_{j=1}^{n_i}\theta_{i,j}$, and compute the final fidelity $\left|\langle C_\alpha^+ \mid C_{\alpha+\theta}^+ \rangle\right|^2$. We repeat this procedure 10,000 times and compute an average fidelity as a function of $N$.

While the Monte-Carlo simulation should be the most accurate method of predicting the fidelity decay curve, we would also like to have a mechanism for assessing the relative importance of each error channel in determining the final dephasing rate. To do so, we first calculate an effective "dephasing per occurrence" for each event, which is a number between 0 and 1 indicating the degree of dephasing induced by the error. The dephasing probability is then the product of the probability of occurrence and dephasing per occurrence (see Table S3). We compute the infidelity *f* per occurrence as follows:

$$
\begin{aligned}
f(\delta\chi, t_0, t_1) &= 1 - \frac{1}{t_1 - t_0} \int_{t_0}^{t_1} \left| \left\langle C_\alpha^+ \mid C_{\alpha+\delta\chi t}^+ \right\rangle \right|^2 dt \\
&= 1 - \frac{4}{(t_1 - t_0)\mathrm{N}^2} \int_{t_0}^{t_1} \left| \left\langle \alpha \mid \alpha + \delta\chi t \right\rangle + \left\langle \alpha \mid \alpha + \delta\chi t + \pi \right\rangle \right|^2 dt \\
&= 1 - \frac{4}{(t_1 - t_0)\mathrm{N}^2} \int_{t_0}^{t_1} \left| e^{-\alpha^2\left(1+\exp\left(-i\delta\chi t\right)\right)} + e^{-\alpha^2\left(1-\exp\left(-i\delta\chi t\right)\right)} \right|^2 dt
\end{aligned}
$$

Where $\left|C_\alpha^+\right\rangle = \left(\left|\alpha\right\rangle + \left|-\alpha\right\rangle\right)/\sqrt{\mathrm{N}}$ , and $\mathrm{N} = 2\left(1 + e^{-2|\alpha|^2}\right)$ is a normalization factor. The effective dephasing per occurrence can be found by comparing the fidelity to the fidelity of the completely dephased state:

$$
\tilde{f}(\delta\chi, t_0, t_1) = \min\left(1, f(\delta\chi, t_0, t_1) / f(\delta\chi, 0, 2\pi/\delta\chi)\right)
$$

## 9. Error Budget

| | Failure mode | | Probability of occurrence | $\delta\chi$ | $[t_0, t_1]$ | Dephasing per occurrence | Probability of dephasing |
|---|---|---|---|---|---|---|---|
| Parity Map | $\lvert f\rangle \to \lvert e\rangle$ | $\Pi_{gf}$ | $\frac{t_{\text{map}}}{2T_1^{fe}}$=4.77% | $\chi_{ef}$ | $[0, t_{\text{map}}]$ | 62% | 2.84% |
| | | $\Pi_{\text{FT}}$ | | 0 | | 0% | 0% |
| | $\lvert f\rangle \to \lvert e\rangle \to \lvert g\rangle$ | | $\frac{1}{4}\frac{t_{\text{map}}^2}{T_1^{fe}T_1^{eg}}$=0.20% | $\chi_{gf}$ | $[\frac{t_{\text{map}}}{3}, t_{\text{map}}]$ | 100% | 0.19% |
| | $\lvert f\rangle \to \lvert h\rangle$ | | $\frac{3}{2}\frac{t_{\text{map}}\bar{n}_{\text{th}}^e}{T_1^{eg}}$=0.38% | $\chi_{fh}$ | $[0, t_{\text{map}}]$ | 100% | 0.38% |
| | $\lvert g\rangle \to \lvert e\rangle$ | | $\frac{1}{2}\frac{t_{\text{map}}\bar{n}_{\text{th}}^e}{T_1^{eg}}$=0.13% | $\chi_{ge}$ | $[0, t_{\text{map}}]$ | 83% | 0.11% |
| Readout | $\lvert g\rangle \to \lvert e\rangle$ | | $\frac{p_g \bar{n}_{\text{th}}^e t_{\text{RO}}}{T_1^{eg}}$=0.12% | $\chi_{ge}$ | $[0, t_{\text{RO}}]$ | 42% | 0.05% |
| | $\lvert e\rangle \to \lvert g\rangle$ | | $\frac{p_e t_{\text{RO}}}{T_1^{eg}}$=0.58% | $\chi_{ge}$ | $[0, t_{\text{RO}}]$ | 42% | 0.25% |
| | $\lvert f\rangle \to \lvert e\rangle$ | | $\frac{p_f t_{\text{RO}}}{T_1^{fe}}$=0.42% | $\chi_{ge}$ | $[0, t_{\text{RO}}]$ | 72% | 0.3% |
| Assignment | Assign $\lvert g\rangle$ as $\lvert e\rangle$ | | 0.04% | $\chi_{ge}$ | $[t_{\text{RO}}, t_{\text{RO}}]$ | 100% | 0.04% |
| | Assign $\lvert e\rangle$ as $\lvert g\rangle$ | | 0.01% | $\chi_{ge}$ | $[t_{\text{RO}}, t_{\text{RO}}]$ | 100% | 0.01% |
| | Assign $\lvert e\rangle$ as $\lvert f\rangle$ | | 0.02% | $\chi_{ef}$ | $[t_{\text{RO}}, t_{\text{RO}}]$ | 100% | 0.02% |
| | Assign $\lvert f\rangle$ as $\lvert e\rangle$ | | 0.01% | $\chi_{ef}$ | $[t_{\text{RO}}, t_{\text{RO}}]$ | 100% | 0.01% |
| | | | | | | Total error probability ($\Pi_{\text{FT}}$) | 1.36% |
| | | | | | | Total error probability ($\Pi_{gf}$) | 4.20% |

**Table S3: Dominant logical errors caused by higher-order ancilla errors.** In these formulae, $t_{\text{map}} = \pi / \chi_{fg}$ = 2.1 μs is the time required to perform the parity mapping, $t_{\text{RO}}$ = 1.2 μs is the time

required to perform the readout of the ancilla. $\{p_g, p_e, p_f\} \approx \{0.8, 0.12, 0.08\}$ are the probabilities of ending the protocol in $g, e, f$, respectively. The probability of ancilla assignment error is estimated from the overlap of the Gaussian distributions in the histograms of readout outcomes, as well as the prior probability of measuring a given state. Dephasing per occurrence is calculated from $\tilde{f}(\delta\chi, t_0, t_1)$ as defined in supplementary material 8.

**10. System Parameters**

| Description | Hamiltonian/Lindblad Term | Measured value |
|---|---|---|
| Cavity Frequency | $\omega_c \hat{a}^\dagger \hat{a}$ . | $\omega_c \approx 2\pi \times 4.5$ GHz |
| Transmon g-e frequency | $\omega_q \lvert e\rangle\langle e\rvert$ . | $\omega_q \approx 2\pi \times 6.5$ GHz |
| Readout resonator frequency | $\omega_{RO} \hat{r}^\dagger \hat{r}$ | $\omega_{RO} \approx 2\pi \times 9.3$ GHz |
| Transmon anharmonicity | $(2\omega_q + \alpha) \lvert f\rangle\langle f\rvert$ | $\alpha \approx 2\pi \times -210$ MHz |
| Cavity Self-Kerr | $\frac{K}{2} \hat{a}^\dagger \hat{a}^\dagger \hat{a}\hat{a}$ | $K \approx 2\pi \times -10$ Hz |
| Transmon-Readout Cross-Kerr | $\chi_{RO} \hat{r}^\dagger \hat{r} \lvert e\rangle\langle e\rvert$ | $\chi_{RO} \approx 2\pi \times -1.3$ MHz |
| Transmon-Cavity Cross-Kerr | $\left(\chi_e \lvert e\rangle\langle e\rvert + \chi_f \lvert f\rangle\langle f\rvert\right) \hat{a}^\dagger \hat{a}$ | $\chi_e = 2\pi \times -93$ kHz,<br>$\chi_f = 2\pi \times -236$ kHz |
| Cavity-Readout Cross-Kerr | $\chi_{ra} \hat{r}^\dagger \hat{r} \hat{a}^\dagger \hat{a}$ | $\chi_{ra} \approx 2\pi \times -0.4$ kHz |
| Cavity Photon Loss | $\frac{1}{T_1^c} D[\hat{a}]$ | $T_1^c \approx 1.07$ ms |
| Transmon $\lvert e\rangle$ to $\lvert g\rangle$ decay | $\frac{1}{T_1^{eg}} D[\lvert g\rangle\langle e\rvert]$ | $T_1^{eg} \approx 25$ μs |
| Transmon $\|f\rangle$ to $\|e\rangle$ decay | $\frac{1}{T_1^{fe}} D[\|e\rangle\langle f\|]$ | $T_1^{fe} \approx 23\,\mu s$ |
| Transmon $\|g\rangle$ dephasing time | $\frac{1}{T_\phi^{g}} D[\|g\rangle\langle g\|]$ | $T_\phi^{g} \approx 81\,\mu s$ |
| Transmon $\|e\rangle$ dephasing time | $\frac{1}{T_\phi^{e}} D[\|e\rangle\langle e\|]$ | $T_\phi^{e} \approx 17\ \ \mu s$ |
| Transmon $\|f\rangle$ dephasing time | $\frac{1}{T_\phi^{f}} D[\|f\rangle\langle f\|]$ | $T_\phi^{f} \approx 12\,\mu s$ |
| Transmon thermal population | $\frac{\bar{n}_{\text{th}}^{e}}{T_1^{eg}} D[\|e\rangle\langle g\|]$ | $\bar{n}_{\text{th}}^{e} \approx 0.02-0.03$ |

**Table S4. System Parameters**